\documentclass[%
 pra, twocolumn,
 superscriptaddress,
 showpacs,
 longbibliography,
 aps,
 showkeys]{revtex4-2}

\usepackage[pdftex]{hyperref,color,graphicx}
\usepackage{amsfonts,amssymb,amsmath}
\usepackage[separate-uncertainty=false]{siunitx}
\usepackage[english]{babel}
\usepackage[utf8]{inputenc}
\usepackage[T1]{fontenc}
\usepackage[toc,page]{appendix}

\usepackage{graphicx}
\usepackage[space]{grffile}
\usepackage{placeins}
\usepackage{latexsym}
\usepackage{textcomp}
\usepackage{longtable}
\usepackage{multirow,booktabs}
\usepackage{url}
\hypersetup{colorlinks=false,pdfborder={0 0 0}}
\usepackage[english]{babel}
\usepackage{dcolumn}
\usepackage{bm}
\usepackage{units}
\usepackage{upgreek}
\usepackage{color}
\usepackage{cancel}
\usepackage{hyperref}
\usepackage{braket}

\newcommand{\bbN}{\mathbb{N}}

\newcommand{\red}[1]{\textcolor{black}{#1}}

\begin{document}

\title{Detecting Topological phase transitions in a double kicked quantum rotor}

\author{Nikolai Bolik}
\affiliation{Institute of Theoretical Physics, Heidelberg University, Philosophenweg 16, 69120 Heidelberg, Germany}

\author{Caspar Groiseau}
\affiliation{Departamento de F\'isica Teórica de la Materia Condensada and Condensed Matter Physics Center (IFIMAC), Universidad Autónoma de Madrid, 28049 Madrid, Spain}

\author{Jerry H. Clark}
\affiliation{Department of Physics, Oklahoma State University, Stillwater, Oklahoma 74078-3072, USA}

\author{Gil S. Summy}
\email{gil.summy1@gmail.com}
\affiliation{Department of Physics, Oklahoma State University, Stillwater, Oklahoma 74078-3072, USA}
\affiliation{Airy3D, 5445 Avenue de Gasp\'e Suite 230, Montr\'eal, Qu\'ebec H2T 3B2, Canada}

\author{Yingmei Liu}
\email{yingmei.liu@okstate.edu}
\affiliation{Department of Physics, Oklahoma State University, Stillwater, Oklahoma 74078-3072, USA}

\author{Sandro Wimberger}
\email{sandromarcel.wimberger@unipr.it}
\affiliation{Dipartimento di Scienze Matematiche, Fisiche e Informatiche, Universit\`{a} di Parma, Parco Area delle Scienze 7/A, 43124 Parma, Italy}
\affiliation{INFN, Sezione di Milano Bicocca, Gruppo Collegato di Parma, Parco Area delle Scienze 7/A, 43124 Parma, Italy}

\begin{abstract}
We present a concrete theoretical proposal for detecting topological phase transitions in double kicked atom-optics kicked rotors with internal spin-1/2 degree of freedom. The implementation utilizes a kicked Bose-Einstein condensate evolving in one-dimensional momentum space. To reduce influence of atom loss and phase decoherence we aim to keep experimental durations short while maintaining a resonant experimental protocol. Experimental limitations induced by phase noise, quasimomentum distributions, symmetries, and the AC-Stark shift are considered. Our results thus suggest a feasible and optimized procedure for observing topological phase transitions in quantum kicked rotors.
\end{abstract}


\maketitle

\section{Introduction}
\label{sec-intro}

Studies of topological phases have found many applications, including revealing topologically protected edge states in topological insulators \cite{PhysRevB.90.195419,qi2011topological,hasan2010colloquium,Kosterlitz_1973,mermin1979topological}. Stabilities of topological invariants are translated upon these edge states, enabling them to be stable against a great variety of perturbations. This robustness to decoherence makes topological phenomena intriguing for many potential applications, e.g., in quantum computing \cite{Kitaev2009} and quantum walks \cite{kitagawa2011topological}.

Topological effects can be simulated by periodically driven systems in a well-controlled manner~\cite{PhysRevLett.65.3076, kitagawa2011topological, Eckardt2017}. A double kicked quantum rotor (DKQR) with internal spin-1/2 degree of freedom is an example of such systems and therefore a potential candidate to experimentally realize topological phase transitions \cite{Gong2018, condmat4010010}.  In the DKQR system a quantity known as a winding number $\nu$ is topologically invariant under a wide range of transformations \cite{Gong2018}. A similar phenomenon occurs in solid state systems where holes are preserved under certain transformations due to geometrical topology~\cite{cayssol2021topological}.   
Preservation of topological winding numbers requires preservation of chiral symmetry and the band gap~\cite{Gong2018, condmat4010010}. A consequence of permanently containing chiral symmetry is that the topological invariant can only change when the system is changed to a configuration in which the band gap closes. This closure makes the phase undetermined but also allows for its direct experimental control by scanning a system parameter through it. 
The same is true for periodically driven systems with Floquet spectra, as in our case of the DKQR \cite{Gong2018}. Here, the quasienergy spectrum itself is periodic and gaps can be controlled by the driving parameters. Floquet topological states were observed in different experimental settings, including ultracold atom \cite{Flo-3, Flo-4}, photonic \cite{MW-Top, Flo-Top, Top-Walk}, phononic and acoustic systems \cite{Mech-Top, Sound-Top, Acoustic-Top}. To understand which topological phases the system has, one examines its spectral symmetries and the related "protected gaps" in the quasieneregy spectrum of the Floquet Hamiltonian, see e.g. \cite{Asb-1, Asb-2, kitagawa2011topological}. In our DKQR systems, the gaps and the phases can be tuned by the kicking strengths, as described in detail in \cite{Gong2018}.

The experimental setup under consideration as described in \cite{dadras2019experimental} consists of a Bose-Einstein condensate (BEC) with two Zeeman hyperfine states $|1\rangle$ and $|2\rangle$ participating in the dynamics, effectively forming a spin-$1/2$ system. DKQR is based on a singly quantum kicked rotor (QKR) \cite{Raizen1999, SW2011} and is described by the Hamiltonian 
\begin{equation}
\begin{split}
    \hat{\mathcal{H}} &= \frac{\hat{p}^2 \otimes \mathbf{1}}{2}+k_1~\mathrm{cos}(\hat{\theta})\otimes\hat{\sigma}_x\sum_{n=0}^\infty\delta(t-2n\tau) \\\
    &+ k_2~\mathrm{sin}(\hat{\theta})\otimes \hat{\sigma}_y \sum_{n=0}^\infty \delta(t-(2n+1)\tau).
\end{split}
\label{eq-ham}
\end{equation}
Here, $\hat{p}$ and $\hat{\theta}$ are respectively the momentum and angular position operators, $\tau$ describes the duration between two kicks of different kicking strengths $k_{1}$ and $k_{2}$, and the Pauli-matrices $\hat{\sigma}_x$ and $\hat{\sigma}_y$ act on internal spin-$1/2$ degree of freedom. 
Under the on-resonance condition of $\tau=4\pi$ \cite{PhysRevLett.96.160403, PhysRevLett.105.054103, PhysRevLett.75.4598,PhysRevLett.98.083004, PhysRevE.83.046218, PhysRevLett.100.024103, PhysRevLett.99.043002}, corresponding to a full revival (at the Talbot time) of the free evolution of the momentum degree of freedom, the quasiperiodicity of the system can lead to a Hofstadter butterfly-like quasienergy spectrum \cite{PhysRevB.14.2239, Bloch2013, Ketterle2013}, resolving a band structure rich in displaying topological properties \cite{Gong2018}.

The previously proposed experimental sequences for achieving topological phase transitions in DKQRs include a sequence of resonant microwave (MW) and standing-wave kicking laser pulses~\cite{Gong2018,condmat4010010}. Similar sequences have been successfully conducted on QKR systems, including ours consisting of $^{87}$Rb BECs \cite{Dadras2018, dadras2019experimental, Clark2021}. For observing the predicted topological effects, the experimental procedure needs to be carefully designed to overcome a number of experimental challenges. In particular, phase noise arising from random phase fluctuations in the MW pulses and a finite quasimomentum distribution of the BEC must be considered. In this paper we demonstrate how to transform the proposals for detecting topological phases in DKQRs \cite{Gong2018, condmat4010010} into a feasible experimental procedure by presenting solutions to some experimental limitations/challenges that appear in the BEC-based quantum walk and DKQR setups \cite{Dadras2018, dadras2019experimental, Clark2021}.

\section{Previous Theoretical Proposals}
\label{sec-oldprotocol}

We briefly review previous proposals for the measurement of the topological phases in the DKQR setup \cite{Gong2018, condmat4010010}. In these proposals, the MW operations correspond to a Rabi coupling between the two internal states $\ket{F=1, m_{F}=0}$ and $\ket{F=2, m_{F}=0}$ of the atoms and are expressed as a unitary rotation on the Bloch sphere:
\begin{equation}
\hat M(\alpha, \chi) = 
 \begin{pmatrix}
\cos(\frac{\alpha}{2})            & e^{-i \chi}\sin(\frac{\alpha}{2})  \\ 
-e^{i \chi}\sin(\frac{\alpha}{2}) & \cos(\frac{\alpha}{2}) \end{pmatrix},
 \label{eq-matrix} 
\end{equation}
The kicking laser is detuned between these internal states in such a way where the potential is equal in strength but opposite in sign, as expressed by a $\hat{\sigma}_z$ matrix. The kick operators, $\hat{K}_1$ and $\hat{K}_2$ differing only by a shift of $\theta=\pi/2$ in position space, are defined as
\begin{align}
    \hat{K}_1 &= e^{-ik_1\mathrm{cos}(\hat{\theta})\hat \sigma_z} \label{eq.k1}\\
    \hat{K}_2 &= e^{-ik_2\mathrm{sin}(\hat{\theta})\hat \sigma_z} \label{eq.k2}.
\end{align}
As discussed in detail in \cite{Gong2018},
the system consists of two chirally symmetric time frames expressed with two Floquet operators, $\hat{U}_1$ and $\hat{U}_2$, possessing chiral symmetry. These operators are best realised as a sequence of MW and kick operators on the atoms' initial wave function, e.g., $\ket{\psi_{\rm in}}=|n=0\rangle\otimes|2\rangle$,
and take the following form:
\begin{align}
\begin{split}
    \hat{U}_1  =&\expandafter\hat M\left( - \frac{\pi}{2} ,0\right)\hat K^{\frac{1}{2}}_1 \expandafter\hat M\left(  \frac{\pi}{2}  ,0\right)\expandafter\hat M\left(-\frac{\pi}{2} ,\frac{\pi}{2}\right)\hat K_2\expandafter\hat M\left(\frac{\pi}{2} ,\frac{\pi}{2}\right)\\
     &\cdot\expandafter\hat M\left( - \frac{\pi}{2} ,0\right)\hat K^{\frac{1}{2}}_1 \expandafter\hat M\left( \frac{\pi}{2}  ,0\right)\label{U1}
\end{split}\\
\begin{split}
   \hat{U}_2  =&  \expandafter\hat M\left(-\frac{\pi}{2} ,\frac{\pi}{2}\right)\hat K^{\frac{1}{2}}_2\expandafter\hat M\left(\frac{\pi}{2} ,\frac{\pi}{2}\right) \expandafter\hat M\left( - \frac{\pi}{2} ,0\right)\hat K_1 \expandafter\hat M\left(  \frac{\pi}{2} ,0\right)\\
   &\cdot\expandafter\hat M\left(-\frac{\pi}{2} ,\frac{\pi}{2}\right)\hat K^{\frac{1}{2}}_2\expandafter\hat M\left(\frac{\pi}{2} ,\frac{\pi}{2}\right)\label{U2}
\end{split}
\end{align}
with
\begin{equation}
    \hat{K}_{1,2}^{1/2} \equiv \hat{K}_{1,2}\left(\frac{k}{2},\theta\right).     
\end{equation}

Due to the structural similarity of both operators we will further focus on the discussion of $\hat{U}_2$, referred to hereafter as $\hat{\mathcal{U}}$ in this paper. Analogous reasoning can be found for $\hat{U}_1$, but is not explicitly shown here.

As an abstract quantity, the topological winding number is not often directly measurable. Instead a quantity, the mean chiral displacement (MCD), is introduced in this context \cite{Gong2018}. In DKQRs the MCD, describing the difference between momentum distributions of the two internal states that evolve under $\hat{\mathcal{U}}$, is defined as:
\begin{equation}
\begin{split}
    C(t) &= \langle \psi_{t} |\hat{n} \otimes - \hat \sigma_z|\psi_t\rangle \\\
    &\equiv \langle \psi_{0} | \hat{\mathcal{U}}^{-t}(\hat{n} \otimes - \hat \sigma_z)\hat{\mathcal{U}}^t|\psi_0\rangle.\label{eq.MCD}
\end{split}
\end{equation}
The average of the MCD over several discrete evolution steps $t$ 
converges to half of the topological winding number $\nu$ \cite{Gong2018}: 
\begin{equation}
        \bar{C}(t) = \frac{1}{t} \sum_{t_i=1}^t C(t_i) 
        \xrightarrow{t \gg 1} \frac{\nu}{2}
        \label{eq.convergence}.
\end{equation}
To observe the topological phase transitions it is necessary to repeat application of Eq.~\eqref{U2} for a series of configurations of $k_1$ and $k_2$ where the empirical results can be compared to the ideal phase diagram computed in \cite{Gong2018}.

\section{Optimized experimental Sequence}
\label{sec-new_protocol}

The first and last MW rotations within Eq. \eqref{U2} are inverses of each other. If $\hat{\mathcal{U}}$ is applied subsequently for a larger number of evolution steps $t \in \bbN$, this sequence of MW rotations and kicks can be simplified as demonstrated in full detail in App. \ref{sec.simplification}.
Considering the full evolution of the system an alternative expression of $\hat{\mathcal{U}}^t$ is therefore found as:
\begin{equation}
\label{eq.U2_new}
\begin{split}
\hat{\mathcal{U}}^t =& \hat{M}(-\frac{\pi}{2},\frac{\pi}{2})\hat{K}_2^{\frac{1}{2}}\hat{M}(\frac{\pi}{2},\frac{\pi}{2}) \\\ &\cdot\left[\hat{M}(-\frac{\pi}{2},0)\hat{K}_1\hat{M}(\frac{\pi}{2},0)\right.\left.\hat{M}(-\frac{\pi}{2},\frac{\pi}{2})\hat{K}_2\hat{M}(\frac{\pi}{2},\frac{\pi}{2}) \right]^{t-1} \\\
& \cdot \hat{M}(-\frac{\pi}{2},0)\hat{K}_1\hat{M}(\frac{\pi}{2},0) \hat{M}(-\frac{\pi}{2},\frac{\pi}{2})\hat{K}_2^{\frac{1}{2}}\hat{M}(\frac{\pi}{2},\frac{\pi}{2}).
\end{split}
\end{equation}
Although this new expression looks more complex, rewriting the sequence in this form significantly reduces the number of operations necessary to realize the complete evolution $\hat{\mathcal{U}}^t$. This is indeed of great interest because it shortens experimental durations, reducing the impact of atom loss and other sources of decoherence (See, e.g., Ref.~\cite{dadras2019experimental} for a short summary of these effects).

So far each application of the MW operator is assumed to be infinitely short in time corresponding to a fully quantum resonant atom-optics kicked rotor \cite{WGF2003, SW2011}.
However, this assumption does not apply to typical experiments since every MW rotation has a finite duration $\mathcal{T}$. A free evolution operator thus needs to be added after each MW operator as follows:
\begin{equation}
    \hat{M}(\alpha,\chi) \rightarrow e^{-i\frac{\hat p^2}{2}\mathcal{T}} \hat{M}(\alpha,\chi).
    \label{eq.addfree_evolution}
\end{equation}

Experimental durations can be further reduced by keeping $\mathcal{T}$ as short as possible. 
When $\mathcal{T}=4\pi$ the free evolution operator reduces to unity and thus reflects a full Talbot period. This in turn corresponds to an on-resonance atom-optics kicked rotor \cite{SW2011, WGF2003}.
In a specific case of Eq. \eqref{eq.U2_new}, it is possible to produce a resonant configuration with $\mathcal{T}=\pi$, corresponding to a good MW signal with a quarter Talbot period.
In this paper, we consider $\mathcal{T}=\pi$ and free evolution of the form  ${e^{-i \frac{\hat{p}^2}{2}\pi} \equiv \hat{P}_\pi}$.
In App.~\ref{prerequisites} the free evolution operator is identified with the shift operator $\hat{T}(\hat{\theta})=e^{i\hat{n}\hat{\theta}}$ where $\theta=\pi$ in position space, thus $\hat{P}_\pi^2=\hat{T}(\pi)$.  As a consequence of introducing the free evolution $\hat{P}_{\pi}$, every second kick operator experiences an effective inversion:
\begin{equation}
    \begin{split}
        \hat{\mathcal{U}}^t &\rightarrow \hat{P}_\pi\hat{M}(-\frac{\pi}{2},\frac{\pi}{2})\hat{K}_2^{\frac{1}{2}}\hat{M}(\frac{\pi}{2},\frac{\pi}{2}) \\\
        &\left[\hat{M}(-\frac{\pi}{2},0)  \hat{K}_1^{-1}\hat{M}(\frac{\pi}{2},0)\right.\left.\cdot\hat{M}(-\frac{\pi}{2},\frac{\pi}{2})\hat{K}_2\hat{M}(\frac{\pi}{2},\frac{\pi}{2}) \right]^{t-1}\\\
        & \cdot \hat{M}(-\frac{\pi}{2},0)\hat{K}_1^{-1}\hat{M}(\frac{\pi}{2},0)\hat{M}(-\frac{\pi}{2},\frac{\pi}{2})\hat{K}_2^{\frac{1}{2}}\hat{P}_\pi\hat{M}(-\frac{\pi}{2},\frac{\pi}{2}).
    \end{split}
    \label{eq.U2_invert}
\end{equation}
More calculation details can be found in App.~\ref{free_evolution}. Although this free evolution effect is undesired, it fortunately does not affect the measurement of topological phase transitions as discussed in Sec.~\ref{sec-symmetry_inversion}.
Note that this induced inversion symmetry (see Eq.~\eqref{eq.U2_invert}) simplifies the observation of topological phase transitions in our $^{87}$Rb system where the required $\pi$ pulse has a duration of $\approx 26 \mu s$ achievable via a high-power MW amplifier~\cite{Alberti2017,Dadras2018, dadras2019experimental, Clark2021}. If $\hat{\mathcal{U}}^t$ is computed for a MW rotation of $\mathcal{T}=\pi$ without the simplification using Eq.~\eqref{eq.U2_new}, the resultant operator sequence is antiresonant inducing periodic oscillations in momentum space. This is understood analytically and verified with numerical calculations. Therefore, a combination of our proposal (see Eqs.~\eqref{eq.U2_new} and \eqref{eq.U2_invert}) and setting $\mathcal{T}=\pi$ is needed for an optimized experimental sequence for realizing topological phase transitions in DKQRs.


\subsection{Inversion symmetry}
\label{sec-symmetry_inversion}

\begin{figure}[tb]
    \centering
    \includegraphics[width=.95\linewidth]{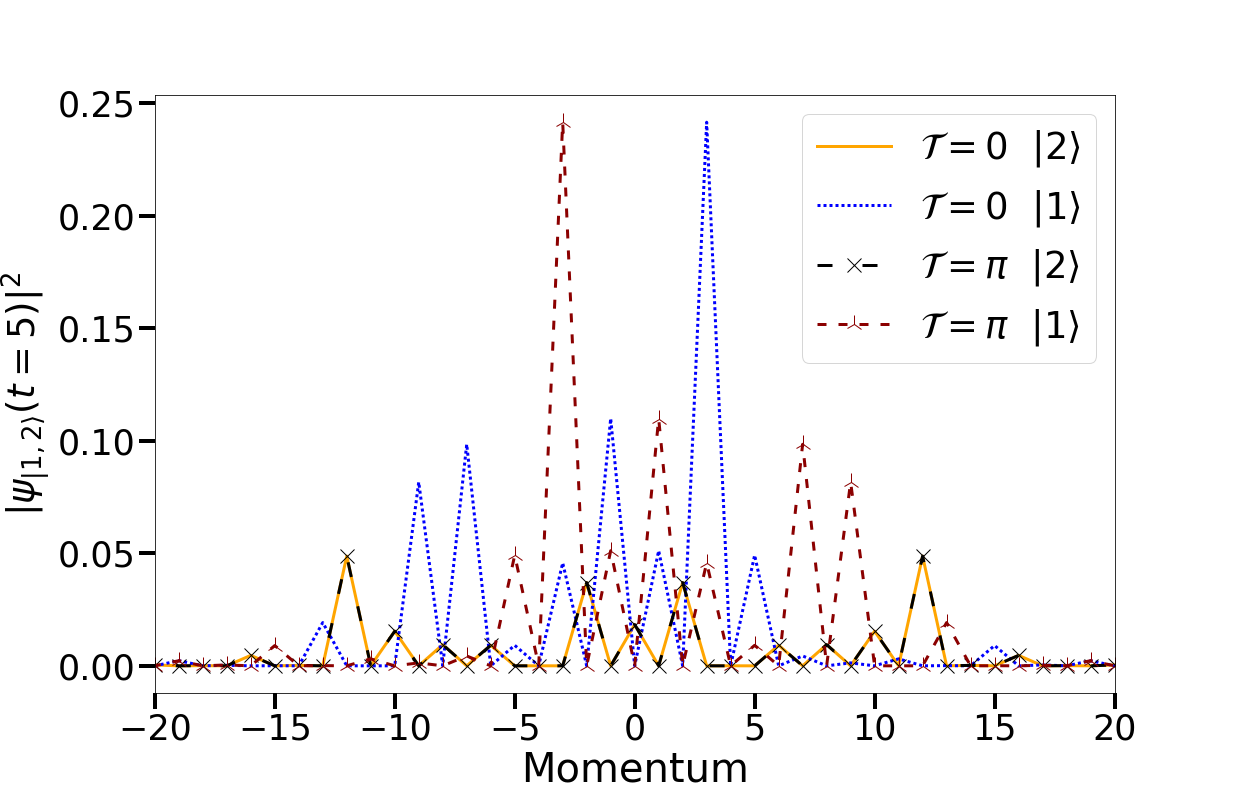}
    \caption{Momentum distributions of the internal states after $t=5$ applications of $\hat{\mathcal{U}}$ for $k_1=\frac{\pi}{2}$ and ${k_2=2.5\pi}$. 
    The initial state is $|\psi_{\rm in}\rangle =|n=0\rangle  \otimes |2\rangle$. $\mathcal{T} =0$ corresponds to fully resonant conditions, while $\mathcal{T} = \pi$ specifies MW pulses with the finite duration. If $\mathcal{T}=\pi$ is chosen, the momentum distributions of the internal states are mirrored respectively to the ideal case. Note that the initial state $\ket{2}$ evolves symmetrically in momentum space thus not contributing to the MCD. }
    \label{fig1}
\end{figure}

A finite duration $\mathcal{T}=\pi$ of each MW rotation results in every second kick operator to be effectively inverted (see App.~\ref{Appendix_3}), i.e.,
\begin{align}
    \hat{K}_1 &= e^{-ik_1\mathrm{cos}(\hat{\theta})\hat\sigma_z} \rightarrow e^{+ik_1\mathrm{cos}(\hat{\theta})\hat \sigma_z} \label{eq.k1b}\\
    \hat{K}_2 &= e^{-ik_2\mathrm{sin}(\hat{\theta})\hat\sigma_z}\rightarrow e^{-ik_2\mathrm{sin}(\hat{\theta})\hat \sigma_z}\label{eq.k2b}.
\end{align}
Instead of a relative shift of $\theta=\pi/2$ in position space in between the two kicks, see
Eqs.~\eqref{eq.k1} and \eqref{eq.k2}, Eqs.~\eqref{eq.k1b} and \eqref{eq.k2b} indicate that the relative shift is $\theta=-\pi/2$. This is equivalent to an inversion in position/angle space around the zero angle. As a result, momentum distributions of the internal states evolve along opposite directions. This causes the expectation value of momentum for each internal state respectively to change sign, as illustrated in Fig.~\ref{fig1}.

Eqs.~\eqref{eq.MCD} and Eq.~\eqref{eq.convergence} also indicate that the topological curves computed from the average MCD change their signs due to introducing the finite MW duration of $\mathcal{T}=\pi$. This does not change the underlying physics and thus the expected phase transitions are still visible although with a change in sign. This undesired effect can be compensated by changing several phase angles of the MW operators. For instance, we find that
\begin{equation}
    \begin{split}
        \hat{M}(-\frac{\pi}{2},0)e^{-ik_1\mathrm{cos}(\hat \theta)\hat \sigma_z}\hat{M}(\frac{\pi}{2},0)\\ = \hat{M}(\frac{\pi}{2},0)e^{ik_1\mathrm{cos}(\hat \theta)\hat \sigma_z}\hat{M}(-\frac{\pi}{2},0).
    \end{split}\label{eq.compensation}
\end{equation}
This change in the phases within the MW rotations causes a change in sign in the exponent and thus compensates for the aforementioned change of the two internal states.

\subsection{Initial state dependence}
\label{sec-initial_state}

Momentum distributions are good observables in QKR and quantum walk experiments, as shown in Refs.~\cite{Dadras2018, dadras2019experimental, Clark2021} where the complete momentum distribution ${P(t) = P(t)_{|1\rangle}+P(t)_{|2\rangle}}$ is measured over one experimental cycle. The momentum distributions for both internal states, $P(t)_{|1\rangle}$ and $P(t)_{|2\rangle}$, however, need to be measured separately to compute the MCD using Eq.~\eqref{eq.MCD}. This may double the experimental effort, because one MCD measurement requires two consecutive experimental cycles with one cycle addressing one individual state in systems similar to those shown in \cite{SW2011, Raizen1999}.
The presence of additional symmetries can ease this requirement because a state symmetrically evolving in momentum space has a zero mean momentum and no contribution to the MCD. Therefore, if one of the internal states is symmetric in momentum space, the information of the topological winding number can be conveniently extracted from the momentum distributions of the other internal state. As a result this effectively reduces the number of experimental measurements in practice. A good example is illustrated in Fig.~\ref{fig1}.

\section{Stability}
\label{sec-stability}

\begin{figure}[!h]
    \centering
    \includegraphics[width=0.9\linewidth]{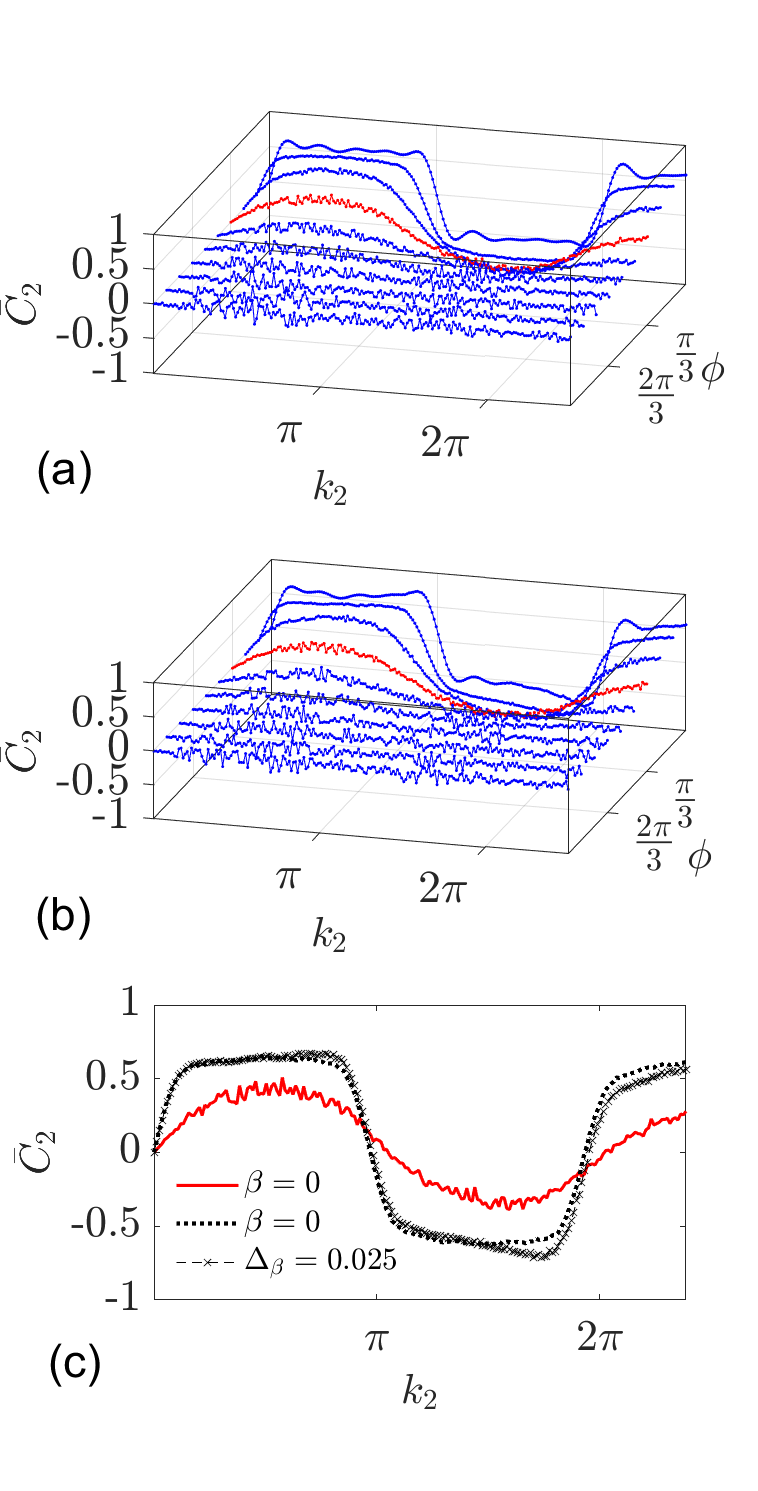}
  \caption{The predicted MCD and topological phase diagrams as a function of the noise strength $\phi$ when $\beta=0$ (a) and $\Delta_\beta=0.025$ (b) derived for our proposed sequence Eq.~\eqref{eq.U2_invert} at $\mathcal{T} = \pi$. The change in sign is compensated using Eq. \eqref{eq.compensation} and the MCD is computed exploiting the symmetry of the initial state $|\psi_{\rm in}\rangle =|n=0\rangle  \otimes |2\rangle$ as discussed in Sec.~\ref{sec-initial_state}. The red/gray line corresponds to $\phi = \pi/3$, the maximum phase noise acceptable for observing topological phase transitions. \red{(c) The black dotted and black dashed lines represent cuts at $\phi=\pi/9$ respectively from panel (a) and (b), and the red/gray solid line corresponds to a cut at $\phi=\pi/3$ from panel (a). While the steps are visible at small noise strength $\phi$ (see the black lines), they get washed out as $\phi$ becomes larger than $\pi/3$.
} For each $k_2$, the topological number is averaged over 1000 noise trajectories in all cases (see text).
}
  \label{fig_2}
\end{figure}
Some experimental imperfections, for example fluctuations in kicking strengths and pulse durations, have little effects on the MCD measurements~\cite{Gong2018} because the kicks can be timed with good precision and resonant atom-optics kicked rotors are intrinsically stable with respect to such imperfections~\cite{SW2011}.
In this section we focus on three unavoidable experimental limitations impacting phase evolutions of the system. First, we will investigate the most crucial problem of uncontrolled phases of the MW pulses. There are many MW pulses during a single experimental cycle (see Eq. \eqref{eq.U2_invert}), thus these phase errors would accumulate quickly. Second, different quasimomenta result in coherent but unwanted deviations from quantum resonance in atom-optics kicked rotors. It is necessary to verify whether the typical spread in momentum reported previously \cite{dadras2019experimental, Dadras2018, Clark2021} has some impact on the measurement of topological phases. The third problem arises from a relative energy shift between the two internal states. This leads to a relative dynamical phase in the evolution that must be corrected.

\subsection{Phase noise}
\label{sec-phase_noise}

One major source of perturbation originates from fluctuations on the precise timing of the internal MW rotations across the ensemble \cite{dadras2019experimental}. This effectively leads to fluctuations on the phase parameter $\chi$ in the MW rotation operator $\hat{M}(\alpha,\chi)$ (see Eq. \eqref{eq-matrix}) as follows
\begin{equation}
    \hat{M}(\alpha,\chi) \rightarrow \hat{M}(\alpha,\chi + \Delta_{\chi}),
    \label{eq.MW_noise}
\end{equation}
modeled numerically as a time-dependent random walk within an additional dynamical phase $\Delta_{\chi} = \sum_{i=1}^{l}\delta_{\chi, i}$. Here ${\delta_{\chi, i} \in \mathrm{uniform}[-\phi, \phi]}$ is drawn individually for each MW application, and $l$ counts the number of subsequent MW applications. 

DKQRs are much more complicated than a single kicked atom-optics kicked rotor (see \cite{SW2011, Raizen1999}), however, our experimental system is capable of generating DKQRs with $t\approx 5$ steps \cite{dadras2019experimental, Dadras2018, Clark2021}. Figure~\ref{fig_2} shows our numerical results of the MCD using Eq.~\eqref{eq.MCD} with $1000$ noise trajectories, and topological diagrams using Eq.~\eqref{eq.convergence} with a time average over up to $t=5$ applications of $\hat{\mathcal{U}}$.
Similar to \cite{condmat4010010}, our simulations keep the kick strength $k_1$ at $\pi/2$ while scanning $k_2$ within a range of $k_2\in [0,\ 2.5\pi]$. Here both kick strengths are expressed in dimensionless units. Scanning $k_2$ reveals topological phase transitions for each configuration of the kick strengths where the band gap \rm{in the Floquet spectrum} closes \cite{Gong2018}. A typical example is shown in Fig.~\ref{fig_2}, which indicates that the signature of the topological phase transitions decays continuously as $\phi$ increases but remains visible for $\phi \le \pi/3$ (see the red/gray curves in Fig.~\ref{fig_2}). Therefore, the MW phase noise should be kept within $\phi<\pi/3$ in experiments.

\subsection{Finite quasimomentum distributions}
\label{sec_quasimomentum}

The periodicity of the kicking potential enables the use of Bloch's theorem with the momentum expressed as $p=n+\beta$. Here $n$ is an integer and $\beta$ is the dimensionless quasimomentum. Up to this point, we have exclusively discussed the case of $\mathcal{T}=\pi$ and $\beta=0$, however, experiments with BECs usually start at a momentum close to $n=0$ with a finite width $\Delta_\beta$ in the Brillouin zone. For example, $\Delta_\beta$ is found to be $0.025$ in Refs.~\cite{dadras2019experimental, Dadras2018, Clark2021}. To model this finite quasimomentum distribution, a Gaussian distribution with zero mean value and FWHM $\Delta_\beta=0.025$ is used in our calculations.
Figure~\ref{fig_2} shows topological phase transitions as a function of the phase noise $\phi$ when different quasimomentum distributions are considered. Comparisons between Fig.~\ref{fig_2}(a) and Fig.~\ref{fig_2}(b) indicate that the quasimomenta present in experiments have little effect on the topological diagrams.

\subsection{Light-shift compensation}
\label{sec_comp}

Our system has two internal hyperfine states, the kicking potential created by a virtual transition to a third state thus creates an additional energy gap between the two states. In our system, an atom-optics kicked rotor with two internal states, this light (or AC-Stark) shift effectively changes the kicking potential to~\cite{Groiseau2017, Groiseau2018, PRA2022}
\begin{align}
    \hat {\mathcal{K}} \equiv  
    2k\cdot\mathrm{cos}^2(\frac{\hat \theta}{2}) &= k(\mathrm{cos}(\hat \theta)+1),
    \label{eq.lightshift}
\end{align}
where the constant term independent of the angle $\theta$
represents the shift of the internal states. The total energy shift to be corrected for is thus $2k$ for a single kick of the kicking strength $k$ in our atom-optics kicked rotor systems and the quantum walk experiments \cite{dadras2019experimental, Dadras2018, Clark2021}, because the second internal state effectively sees a kick with an opposite sign.

To observe topological phases, it is important to have a relative shift of $\Delta\theta=\pi/2$ in position space between the two kicks $\hat{K}_1$ and $\hat{K}_2$ (see Eqs. \eqref{eq.k1} and \eqref{eq.k2}).
The effective kick operators are then
\begin{align}
    \hat{K}_{1,\rm eff} &= e^{-ik_1\hat \sigma_z(\mathrm{cos}(\hat{\theta})+1)} \equiv e^{-i\hat \sigma_z \hat{\mathcal{K}}_1} \label{eq.K1eff}\\
    \hat{K}_{2,\rm eff} &= e^{-ik_2\hat \sigma_z(\mathrm{sin}(\hat{\theta})+1)}\equiv e^{-i\hat \sigma_z \hat{\mathcal{K}}_2}.
    \label{eq.k2eff}
\end{align}
\begin{figure*}[!th]
    \centering
    \includegraphics[width=\linewidth]{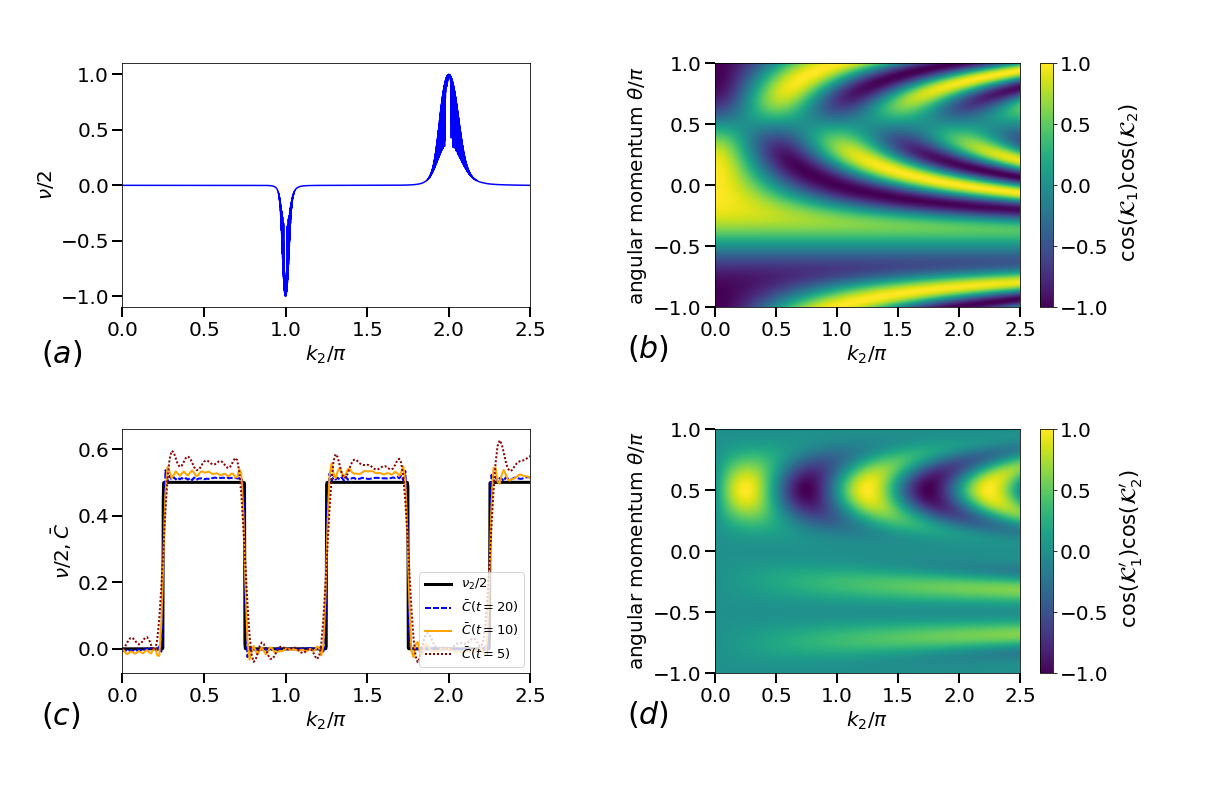}
    \caption{(a) The winding number and (b) values of the function $\mathrm{cos}(\mathcal{K}_1)\mathrm{cos}(\mathcal{K}_2)$ respectively derived from Eqs.~\eqref{eq.K1eff} and \eqref{eq.k2eff} when light-shift effects are NOT compensated. (c) The winding number and (d) values of the function $\mathrm{cos}(\mathcal{K}_1)\mathrm{cos}(\mathcal{K}_2)$ respectively derived from Eqs.~\eqref{eq.K1comp} and \eqref{eq.k2comp} when light-shift effects are effectively compensated. In panel (c), the averaged MCD converges to the predicted phase transitions \red{(black solid line)} as the evolution step $t$ increases. If the $\mathrm{cos}(\mathcal{K}_1)\mathrm{cos}(\mathcal{K}_2)$ function changes sign, the topological phase changes thus implying instabilities. This is satisfied everywhere in panel (b) while only at specific points in panel (d), see the main text. 
    }
    \label{fig_3}
\end{figure*}

The effective kick potentials are described by $\hat{\mathcal{K}}_1$ and $\hat{\mathcal{K}}_2$. No topological phase transitions are observed if these operators are implemented without compensating light shifts as shown in Fig.~\ref{fig_3}(a).
This is because a topological winding number can only change if $\mathrm{cos}(\mathcal{K}_1)\mathrm{cos}(\mathcal{K}_2)=\pm1$ corresponding to a closing band gap (see Ref.~\cite{Gong2018}). Figure~\ref{fig_3}(b) shows that this condition is always fulfilled within the simulated range of $k_2$, thus it becomes impossible to distinguish between topological phase transitions and possible instability, if light shifts are not compensated. This suggests that compensation of light shifts is necessary for observation of topological phase transitions in experiments. 

In walk experiments as reported in Refs.~\cite{dadras2019experimental, Dadras2018, Clark2021}, compensation of light shifts was done by adjusting the phase of MW pulses. Such a procedure is necessary for each kick in the DKQR system where two kicks have different kicking strengths. Our simulations, however, indicate that compensating one of the two kicking pulses is sufficient. For example, a simpler method of compensation can be conducted by correcting light shift originating from $k_1$, therefore the kicking operators are
\begin{align}
    \hat{K}_{1,\rm eff} &  \rightarrow e^{-ik_1\hat \sigma_z\mathrm{cos}(\hat{\theta})} \equiv e^{-i\hat \sigma_z\hat{\mathcal{K^\prime}}_1}  
    \label{eq.K1comp}\\
    \hat{K}_{2, \rm eff} & \rightarrow e^{-ik_2\hat\sigma_z\mathrm{sin}(\hat{\theta})}e^{-i(k_2-k_1)\hat \sigma_z} \equiv e^{-i\hat \sigma_z\hat{\mathcal{K^\prime}}_2}
    \label{eq.k2comp},
\end{align}
where the effective kick dynamics described by $\hat{\mathcal{K}^\prime}_1$ and $\hat{\mathcal{K}^\prime}_2$ are partially compensated. This compensation may not remove light-shift effects completely, however, it is adequate for observing topological phase transitions as shown in Figs.~\ref{fig_3}(c) and~\ref{fig_3}(d). Our calculations show that the number of topological phase transitions doubles in the same range of $k_2$ when the light shift is compensated. This implies that it may be possible to observe topological phase transitions by scanning $k_2$ even in a smaller range after light shifts are effectively compensated.

\section{Conclusion}
\label{sec_con}

We have investigated the feasibility of measuring topological phase transitions with the DKQR platform and demonstrated how impacts of decoherence and experimental durations can be minimized by setting the duration of each MW rotation at $\mathcal{T}=\pi$ corresponding to a quarter of Talbot period. A proper choice of initial states possessing an inversion symmetry is found to further simplify experimental realizations.

Our results have suggested that an successful protocol for observing topological phase transitions must minimize the phase noise to $\phi\leq\pi/3$ and at least partially compensate the light-shift effects. Compensating light-shift effects can also facilitate experiments as it doubles the amount of expected topological steps within a same $k_2$ scanning range. Our findings have thus confirmed that the DKQR is a promising platform to realize and measure topological phases in a time-dependent Floquet setup.

\begin{acknowledgments}
J. H. C. and Y. L. thank the Noble Foundation for financial support.
\end{acknowledgments}

\onecolumngrid
\appendix
\FloatBarrier
\allowdisplaybreaks

\section{Simplification of original proposed sequence}\label{sec.simplification}
\label{Appendix_1
}
For the evolution operator $\hat{\mathcal{U}}$ the first and the last MW-rotation are inverse to each other:
\begin{align}
  \hat M(\frac{\pi}{2},0) = \hat M(-\frac{\pi}{2},0)^{-1}\qquad\qquad\qquad\qquad
  \hat M(-\frac{\pi}{2},\frac{\pi}{2}) = \hat M(\frac{\pi}{2},\frac{\pi}{2})^{-1}.
\end{align}
When $\hat{\mathcal{U}}$ is applied subsequently, this can be exploited to reduce the amount of necessary MWs and kicks in $\hat{\mathcal{U}}$. It is sufficient to study this effect for $\hat{\mathcal{U}}^2$.
\begin{align}
\hat{\mathcal{U}}^2  =&  \left(\expandafter\hat M(-\frac{\pi}{2} ,\frac{\pi}{2})\hat K^{\frac{1}{2}}_2\expandafter\hat M(\frac{\pi}{2} ,\frac{\pi}{2}) \expandafter\hat M( - \frac{\pi}{2} ,0)\hat K_1 \expandafter\hat M(  \frac{\pi}{2} ,0)
\expandafter\hat M(-\frac{\pi}{2} ,\frac{\pi}{2})\hat K^{\frac{1}{2}}_2\expandafter\hat M(\frac{\pi}{2} ,\frac{\pi}{2})\right)^2\\\
=& \expandafter\hat M(-\frac{\pi}{2} ,\frac{\pi}{2})\hat K^{\frac{1}{2}}_2\expandafter\hat M(\frac{\pi}{2} ,\frac{\pi}{2}) \expandafter\hat M( - \frac{\pi}{2} ,0)\hat K_1 \expandafter\hat M(  \frac{\pi}{2} ,0)
\expandafter\hat M(-\frac{\pi}{2} ,\frac{\pi}{2})K^{\frac{1}{2}}_2\\
&\cdot \hat K^{\frac{1}{2}}_2\expandafter\hat M(\frac{\pi}{2} ,\frac{\pi}{2}) \expandafter\hat M( - \frac{\pi}{2} ,0)\hat K_1 \expandafter\hat M(  \frac{\pi}{2} ,0)
\expandafter\hat M(-\frac{\pi}{2} ,\frac{\pi}{2})\hat K^{\frac{1}{2}}_2\expandafter\hat M(\frac{\pi}{2} ,\frac{\pi}{2})\\\
=& \expandafter\hat M(-\frac{\pi}{2} ,\frac{\pi}{2})\hat K^{\frac{1}{2}}_2\expandafter\hat M(\frac{\pi}{2} ,\frac{\pi}{2}) \expandafter\hat M( - \frac{\pi}{2} ,0)\hat K_1 \expandafter\hat M(  \frac{\pi}{2} ,0)\expandafter\hat M(-\frac{\pi}{2} ,\frac{\pi}{2})\hat K_2 \expandafter\hat M(\frac{\pi}{2} ,\frac{\pi}{2})\\
&\cdot  \expandafter\hat M( - \frac{\pi}{2} ,0)\hat K_1 \expandafter\hat M(  \frac{\pi}{2} ,0)
\expandafter\hat M(-\frac{\pi}{2} ,\frac{\pi}{2})\hat K^{\frac{1}{2}}_2\expandafter\hat M(\frac{\pi}{2} ,\frac{\pi}{2})
\end{align}

\section{Conditions necessary for sequence resonance considerations}
\label{prerequisites}
\label{Appendix_2}

We consider the free evolution operator $\expandafter\hat P = e^{-i \frac{\hat p^2}{2} \mathcal{T}}$ for a quarter Talbot time $\mathcal{T} = \pi$. Using Bloch's theorem we can decompose momentum $\hat p$ into an integer and a quasimomentum $\beta$: $\hat p = \hat n + \beta$, with $\beta \in [0,1)$. In the following we choose resonant values for $\beta$, for instance $\beta=0$ \cite{SW2011}, to simplify the discussion. We can identify similarly to \cite{Wagner2018, WGF2003} the free evolution operator with the shift operator in momentum space:
\begin{equation}
\hat P_\pi^2= (e^{-i \frac{ \hat n ^2}{2} \pi}) ^2 = e^{-i  \hat n ^2 \pi} = \left\{ \begin{array}{ll}
1 & n = 2j \\
-1 & n = 2j+1 \\
\end{array} \right. \equiv e^{-i\hat n\pi} \equiv \hat {T}(\pi)  .
\end{equation}
$\hat{T}(\pi)$ can thus denote the shift operator in angular momentum space for $\theta=\pi$. The free evolution operator for the two-state system commutes with the microwave operators since the entries of the MW-matrix are scalar values.
\begin{align}
    \hat{P}= e^{-i \frac{(\expandafter\hat n +\beta)^2}{2} \tau\otimes \mathbf{1}} &=e^{-i \frac{(\expandafter\hat n +\beta)^2}{2} \tau}\cdot \mathbf{1}\\
    \Rightarrow \hat{P}\hat{M}(\alpha_1 , \chi_1)\hat{P}\hat{M}(\alpha_2 , \chi_2)&= \hat{M}(\alpha_1 , \chi_1)\hat{M}(\alpha_2 , \chi_2) \hat{P}^2\\
     &=\hat{M}(\alpha_1 , \chi_1)\hat{M}(\alpha_2 , \chi_2) \hat{T}(\pi).
\end{align}
The translation operator affects the kick operators $\hat{K}_1$ and $\hat{K}_2$ as follows:
\begin{align}
    \hat{T}(\pi)\hat{K}_{1,2}(k,\theta) &= \hat{K}_{1,2}(k,\theta+\pi) = \hat{K}_{1,2}(-k,\theta) \equiv  \hat{K}_{1,2}^{-1}\\
    \hat{T}(2\pi)\hat{K}_{1,2}(k,\theta) &= \hat{K}_{1,2}(k,\theta+2\pi) = \hat{K}_{1,2}(k,\theta).
\end{align}

\section{Introduction of free evolution}\label{free_evolution}
\label{Appendix_3}

The free evolution operator $\hat{P}_\pi$ for $\mathcal{T}=\pi$ after neglecting quasimomentum is not unity. When each MW-rotation is considered to have a finite duration, this affects the evolution of the system described by $\hat{U}^t$ in the following way:
\begin{align}
\begin{split}
    \hat{\mathcal{U}}^t &= \hat{P}_\pi\hat{M}(-\frac{\pi}{2},\frac{\pi}{2})\hat{K}_2^{\frac{1}{2}}\hat{P}_\pi\hat{M}(\frac{\pi}{2},\frac{\pi}{2}) \left[\hat{P}_\pi\hat{M}(-\frac{\pi}{2},0)\hat{K}_1\hat{P}_\pi\hat{M}(\frac{\pi}{2},0) \hat{P}_\pi\hat{M}(-\frac{\pi}{2},\frac{\pi}{2})\hat{K}_2\hat{P}_\pi\hat{M}(\frac{\pi}{2},\frac{\pi}{2})
    \right]^{t-1}\\\
    & \qquad\qquad\qquad\qquad\qquad\qquad \cdot \hat{P}_\pi\hat{M}(-\frac{\pi}{2},0)\hat{K}_1\hat{P}_\pi\hat{M}(\frac{\pi}{2},0)\hat{P}_\pi\hat{M}(-\frac{\pi}{2},\frac{\pi}{2})\hat{K}_2^{\frac{1}{2}}\hat{P}_\pi\hat{M}(\frac{\pi}{2},\frac{\pi}{2})
\end{split}\\
\begin{split}
    &= \hat{P}_\pi\hat{M}(-\frac{\pi}{2},\frac{\pi}{2})\hat{K}_2^{\frac{1}{2}}\hat{M}(\frac{\pi}{2},\frac{\pi}{2}) \left[\hat{M}(-\frac{\pi}{2},0) \hat{T}(\pi) \hat{K}_1\hat{M}(\frac{\pi}{2},0) \hat{M}(-\frac{\pi}{2},\frac{\pi}{2})\hat{T}(\pi)\hat{K}_2\hat{M}(\frac{\pi}{2},\frac{\pi}{2}) \right]^{t-1}\\\
    & \qquad\qquad\qquad\qquad\qquad\qquad\cdot \hat{M}(-\frac{\pi}{2},0)\hat{T}(\pi)\hat{K}_1\hat{M}(\frac{\pi}{2},0)\hat{M}(-\frac{\pi}{2},\frac{\pi}{2})\hat{T}(\pi)\hat{K}_2^{\frac{1}{2}}\hat{P}_\pi\hat{M}(\frac{\pi}{2},\frac{\pi}{2})
\end{split}\\
\begin{split}
     &= \hat{P}_\pi\hat{M}(-\frac{\pi}{2},\frac{\pi}{2})\hat{K}_2^{\frac{1}{2}}\hat{M}(\frac{\pi}{2},\frac{\pi}{2}) \left[\hat{M}(-\frac{\pi}{2},0)  \hat{K}_1^{-1}\hat{M}(\frac{\pi}{2},0) \hat{M}(-\frac{\pi}{2},\frac{\pi}{2})\hat{T}(2\pi)\hat{K}_2\hat{M}(\frac{\pi}{2},\frac{\pi}{2}) \right]^{t-1}\\\
     & \qquad\qquad\qquad\qquad\qquad\qquad\cdot  \hat{M}(-\frac{\pi}{2},0)\hat{K}_1^{-1}\hat{M}(\frac{\pi}{2},0)\hat{M}(-\frac{\pi}{2},\frac{\pi}{2})\hat{T}(2\pi)\hat{K}_2^{\frac{1}{2}}\hat{P}_\pi\hat{M}(\frac{\pi}{2},\frac{\pi}{2})
\end{split}\\
\begin{split}
    & = \hat{P}_\pi\hat{M}(-\frac{\pi}{2},\frac{\pi}{2})\hat{K}_2^{\frac{1}{2}}\hat{M}(\frac{\pi}{2},\frac{\pi}{2}) \left[\hat{M}(-\frac{\pi}{2},0)  \hat{K}_1^{-1}\hat{M}(\frac{\pi}{2},0) \hat{M}(-\frac{\pi}{2},\frac{\pi}{2})\hat{K}_2\hat{M}(\frac{\pi}{2},\frac{\pi}{2}) \right]^{t-1}\\\
    & \qquad\qquad\qquad\qquad\qquad\qquad\cdot \hat{M}(-\frac{\pi}{2},0)\hat{K}_1^{-1}\hat{M}(\frac{\pi}{2},0)\hat{M}(-\frac{\pi}{2},\frac{\pi}{2})\hat{K}_2^{\frac{1}{2}}\hat{P}_\pi\hat{M}(\frac{\pi}{2},\frac{\pi}{2}).
\end{split}
\end{align}
Therefore, every second kick operator (here $\hat{K}_1$) effectively experiences an inversion.

\twocolumngrid

%


\end{document}